\documentclass{iau} 
\usepackage{graphicx}
\usepackage{hyperref}
\usepackage{natbib}

\newcommand*\aap{A\&A}

\newcommand*\apj{ApJ}

\newcommand*\araa{ARA\&A}

\newcommand*\mnras{MNRAS}

\def\arxivprefixesep{:}

\newcommand{\eprint}[2][]{%
{\tt\if!#1!#2\else#1\arxivprefixesep\ignorespaces#2\fi}%
}

\title[Gaia view of low-mass star formation] %% give here short title %%
{Gaia view of low-mass star formation}

\author[C.F. Manara et al.]   %% give here short author list %%
{C.F. Manara$^1$,
\thanks{Present address: European Southern Observatory, Karl-Schwarzschild-Strasse 2, 85748 Garching bei München, Germany, e-mail: cmanara@eso.org.},
 T. Prusti$^1$,
J. Voirin$^1$,
\and E. Zari$^2$
}

\affiliation{$^1$Scientific Support Office, Directorate of Science, European Space Research and Technology Centre (ESA/ESTEC), Keplerlaan 1, 2201 AZ Noordwijk, The Netherlands \\ email: {\tt cmanara@cosmos.esa.int} \\[\affilskip]
$^2$Leiden Observatory, Niels Bohrweg 2, 2333 CA Leiden, the Netherlands}

\pubyear{2015}
\volume{330}  %% insert here IAU Symposium No.
\jname{Astrometry and Astrophysics in the Gaia sky}
\editors{A. Recio-Blanco, P. de Laverny, A. Brown, and T. Prusti editors}
\begin{document}

\maketitle

\begin{abstract}
Understanding how young stars and their circumstellar disks form and evolve is key to explain how planets form. The evolution of the star and the disk is regulated by different processes, both internal to the system or related to their environment. The former include accretion of material onto the central star, wind emission, and photoevaporation of the disk due to high-energy radiation from the central star. These are best studied spectroscopically, and the distance to the star is a key parameter in all these studies. Here we present new estimates of the distance to a complex of nearby star-forming clouds obtained combining TGAS distances with measurement of extinction on the line of sight. Furthermore, we show how we plan to study the effects of the environment on the evolution of disks with Gaia, using a kinematic modelling code we have developed to model young star-forming regions.
\keywords{-}
\end{abstract}

\firstsection % if your document starts with a section,
\section{Introduction}

Protoplanetary disks, made of dust and gas, are formed around young stars and are the birthplace of planets. These disks are evolving with time, and the evolution of their gas and dust content strongly impacts what kind of planets are formed, and their migration in the disk \citep[e.g.,][]{Thommes08,Morby16}. Each of the processes known to impact disk evolution modifies the dust and gas content differently, and it is therefore mandatory to understand which processes are at place during the various disk evolution phases to be able to build a predictive planet formation theory.

The physical processes impacting disk evolution are either internal to the star-disk system, as in the cases of viscous evolution, internal photoevaporation, and disk wind driven evolution \citep[e.g.,][]{Alexander14,Bai16}, or related to the stellar environment where disks form, as in the case of external photoevaporation or dynamical interactions \citep[e.g.,][]{Clarke07,Pfalzner05}. While the former have been studied in details both from a theoretical and observational point of view, the latter are more elusive to be studied observationally, and their impact on disk evolution and planet formation is still to be understood. This is particularly true for dynamical interactions in young clusters and their impact on disk evolution. Pioneering studies have been carried out only in a limited number of environments, mainly in the Orion Nebula Cluster \citep[e.g.,][]{deJuanOvelar}.

The ESA mission Gaia \citep{Gaia} will be of enormous impact in the study of the evolution of young stars and of their disks. In particular, it provides solid estimates of the distance to young stars, a key parameter to derive any stellar and disk property, and it allows one to establish kinematic membership of stars in clusters. Here we present our initial work on these aspects of young stars and disks evolution.

\section{Gaia DR1 distances to the Chameleon clouds}

Distances to nearby star-forming regions are nowadays still uncertain. Even if their distance can be as small as $\sim$50-300 pc, these nearby regions contain mainly highly extincted low-mass stars, which are thus not bright enough to be part of the Hipparcos catalog, the most extensive catalog of parallaxes prior to Gaia. However, the distance to these regions is key to constrain several crucial parameters to study young stars, such as the stellar luminosity and mass, the age, and the disk mass and size.

\begin{figure}[b]
% \vspace*{-2.0 cm}
\begin{center}
 \includegraphics[width=0.7\textwidth]{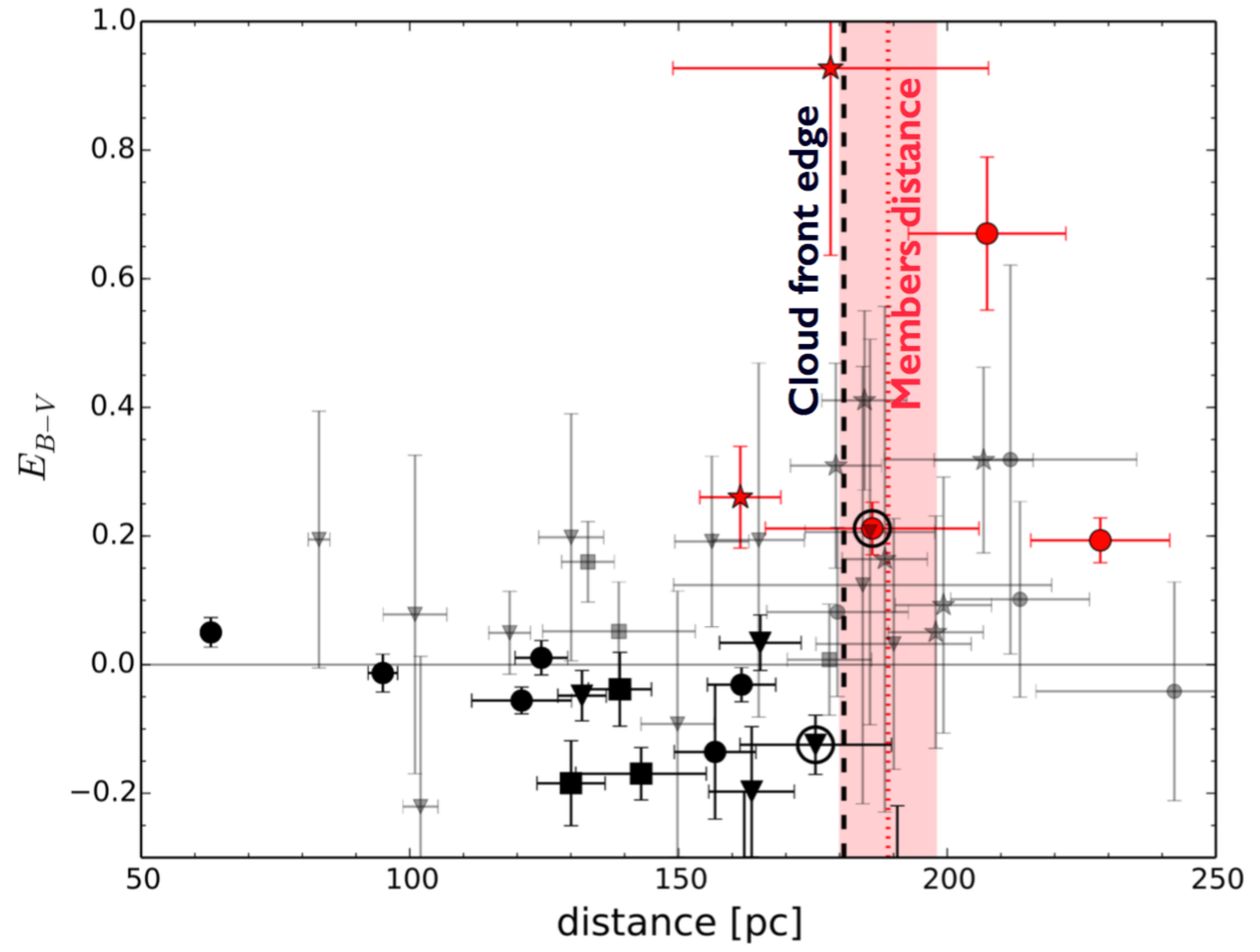} 
% \vspace*{-1.0 cm}
 \caption{Color excess as a function of distance for stars located on the line-of-sight of the Chamaeleon~I cloud. The black dashed line is the position of the front edge of the cloud determined by the reddening turn-on method. The red dotted line is the average distance of the members, shown here for comparison, and the red shaded regions is the uncertainty on this value. 
We highlight with black circles the stars used to compute the distance of the reddening turn-on, thus the front edge of the cloud. From Voirin et al., subm.}
   \label{fig::chaI_dist}
\end{center}
\end{figure}

With the advent of the Gaia catalog, and in particular of the TGAS sub-sample, we have decided to investigate whether we could confirm or update the previous assumed distances to the clouds in one of the closest and most studied star-forming complex, the Chamaeleon-Musca complex \citep[e.g.,][]{LuhmanCha}. This region contains several hundreds of young stars, which are surrounded by disks in many cases \citep[e.g.,][]{Luhman04}, and are still accreting material from these disks \citep[e.g.,][]{Manara16a,Manara17a}. The distances to these clouds have been long debated in the literature and are assumed to be $\sim$160, 178, and 140 pc, for the Chamaeleon~I,~II, and~III clouds, respectively \citep{Whittet97}, while the distance to the Musca cloud is still unknown. 

We have queried the TGAS sub-sample in the Gaia archive for known members of any of these clouds, and we have found 8 members of the Chamaeleon~I cloud and none in the other clouds. Of these 8 stars included in the TGAS catalog, 4 were also included in the Hipparcos catalog. We have compared the distances inferred by converting the parallaxes using an anisotropic prior \citep{Astraatmadja16}, and we observe that the distances obtained from the TGAS parallaxes are systematically larger than the correspondent Hipparcos distances by $\sim$5-25 pc, although the two values are compatible within 1$\sigma$. Furthermore, the TGAS distances to these objects suggest the presence of a north-south gradient in the Chamaeleon~I cloud, with difference of $\sim$20 pc in the distances to the stars, which are found more distant in the southern part of the cloud. When considering these objects, the mean distance to the cloud is 189$\pm$9$\pm$10 pc, where the quoted uncertainties are the statistical and systematic ones, respectively. However, this value is based on a limited number of objects, and the method of computing the distance to a region using the measured distances to its confirmed members cannot be used for the other clouds in this complex at this stage.

Following \citet{Whittet97}, we measure the distance to star-forming clouds by measuring the color-excess due to extinction for objects located on the line-of-sight of the cloud. The presence of the cloud results in an increase of the color-excess, and the distance at which this happens is representative of the distance of the front edge of the cloud. With respect to \citet{Whittet97}, the TGAS catalog gives us access to many more stars with directly measured parallaxes. 
We show in Fig.~\ref{fig::chaI_dist} the measured color-excess vs the distance to the stars on the line-of-sight of the Chamaeleon~I cloud. The distance at which we observe a turn-on in the color-excess is 181$\pm$10$\pm$10 pc, which is compatible with the distance of the members, and is considered as the distance to the region. We similarly derive a distance of 181$\pm$10$\pm$10 pc for the Chamaeleon~II cloud, of 199$\pm$15$\pm$13 pc to the Chamaeleon~III cloud, and we constrain the Musca cloud to be closer than 600 pc (Voirin et al., subm.). We confirm the distance to the Chamaeleon~II cloud, while we find further away distances for the Chamaeleon~I and III clouds, giving for the first time a firm estimate of the distance to the latter. Within the uncertainties, these values are compatible with the hypothesis that all these clouds are part of the same large-scale structure of clouds.

While we have applied this method only to the clouds in the Chamaeleon-Musca complex, this could be applied to any cloud with limited amount of foreground interstellar material. This method is particularly interesting for clouds which contain no members with measured distances, either because of the high extinction or because star formation has not started yet. The larger number of line-of-sight and members stars with measured parallaxes that will be available with DR2 will thus allow us to determine the distance to several other star-forming regions.

\section{Gaia as a tool to study dynamical evolution of young stars}

Young stars are usually located in clustered environment \citep[e.g.,][]{Lada03} where dynamical interactions with other members of the region can affect the evolution of disks \citep[e.g.,][]{Armitage97,Pfalzner15}. Gaia can give us the possibility to probe observationally both the dynamical properties of young clusters \citep[e.g.,][]{Allison12} and the effect of dynamical encounters on disk evolution. Members of star-forming regions are usually identified by the presence of an excess emission at near- and mid-infrared wavelengths due to the presence of a disk. Such method has two main limitations imposed by the small coverage of infrared satellites and by being very little sensitive to objects with small infrared excess, such as disk-less young stars. While the former limitation is now getting overcome by all-sky surveys such as WISE, the latter is possibly biasing our membership lists in young regions. Indeed, surveys of young stars done with other methods, such as measuring H$\alpha$- or UV-excess, or by obtaining spectra of large number of objects in a region, are showing that a large population of young stars is present in the surrounding of the main star-forming site\citep[e.g.,][]{DeMarchi11,Comeron13,Sanchez14,Sacco15}. We think that this sparse population could be either the outcome of dynamical interactions in the main star-forming site, or due to the fact that star formation happens at all degrees of clustering. Both hypotheses are interesting and should be further investigated.

We have been developing a method to assign membership to stars in clusters purely based on their kinematic properties measured by the Gaia satellite. This method is based on the maximum likelihood estimate described by \citet{Lindegren00} and \citet{deBruijne99}, where the membership of a star in a region is assigned by finding the stars sharing a common velocity pattern for which the likelihood is maximized. We have added the possibility to model two populations: a clustered population sharing a common velocity pattern, and a field population (Zari et al., in prep.). With our method we can thus assign a probability to each star to be either member of the cluster or of the field, thus allowing us to identify possible members of a cluster with no assumptions on the presence of a disk or accretion. 

We are now testing this method using N-body simulations of young clusters \citep[e.g.,][]{Parker14} and the GUMS simulation of the stars in the galaxy \citep{GUMS} using the current estimates of the uncertainty on the astrometric parameters in Gaia DR2. The first results show that we assign a probability of being members of the cluster larger than 50\% to $>$95\% of the real members of the cluster, with only $<$5\% of field stars erroneously classified as members.

\acknowledgments{
This work has made use of data from the European Space Agency (ESA)
mission {\it Gaia} (\url{https://www.cosmos.esa.int/gaia}), processed by
the {\it Gaia} Data Processing and Analysis Consortium (DPAC,
\url{https://www.cosmos.esa.int/web/gaia/dpac/consortium}). Funding for
the DPAC has been provided by national institutions, in particular the
institutions participating in the {\it Gaia} Multilateral Agreement.
}


\begin{thebibliography}{}
\bibitem[Alexander et al.(2014)]{Alexander14} Alexander, R., Pascucci, I., Andrews, S., et al.\ 2014, Protostars and Planets VI, 475 

\bibitem[Allison(2012)]{Allison12} Allison, R.~J.\ 2012, \mnras, 421, 3338 

\bibitem[Armitage \& Clarke(1997)]{Armitage97} Armitage, P.~J., \& Clarke, C.~J.\ 1997, \mnras, 285, 540 

\bibitem[Astraatmadja \& Bailer-Jones(2016)]{Astraatmadja16} Astraatmadja, T.~L., \& Bailer-Jones, C.~A.~L.\ 2016, \apj, 833, 119 


\bibitem[Bai(2016)]{Bai16} Bai, X.-N.\ 2016, \apj, 821, 80 

\bibitem[Clarke(2007)]{Clarke07} Clarke, C.~J.\ 2007, \mnras, 376, 1350 

\bibitem[Comer{\'o}n et al.(2013)]{Comeron13} Comer{\'o}n, F., Spezzi, L., L{\'o}pez Mart{\'{\i}}, B., \& Mer{\'{\i}}n, B.\ 2013, \aap, 554, A86 

\bibitem[de Bruijne(1999)]{deBruijne99} de Bruijne, J.~H.~J.\ 1999, \mnras, 310, 585 

\bibitem[de Juan Ovelar et al.(2012)]{deJuanOvelar} de Juan Ovelar, M., Kruijssen, J.~M.~D., Bressert, E., et al.\ 2012, \aap, 546, L1 

\bibitem[De Marchi et al.(2011)]{DeMarchi11} De Marchi, G., Paresce, F., Panagia, N., et al.\ 2011, \apj, 739, 27 

\bibitem[Gaia Collaboration et al.(2016)]{GaiaDR1} Gaia Collaboration, Brown, A.~G.~A., Vallenari, A., et al.\ 2016, \aap, 595, A2 

\bibitem[Gaia Collaboration et al.(2016)]{Gaia} Gaia Collaboration, Prusti, T., de Bruijne, J.~H.~J., et al.\ 2016, \aap, 595, A1 

\bibitem[Lada \& Lada(2003)]{Lada03} Lada, C.~J., \& Lada, E.~A.\ 2003, \araa, 41, 57 

\bibitem[Lindegren et al.(2000)]{Lindegren00} Lindegren, L., Madsen, S., \& Dravins, D.\ 2000, \aap, 356, 1119 

\bibitem[Luhman(2004)]{Luhman04} Luhman, K.~L.\ 2004, \apj, 602, 816 

\bibitem[Luhman(2008)]{LuhmanCha} Luhman, K.~L.\ 2008, Handbook 
of Star Forming Regions, Volume II, 169 

\bibitem[Manara et al.(2016)]{Manara16a} Manara, C.~F., Fedele, D., Herczeg, G.~J., \& Teixeira, P.~S.\ 2016, \aap, 585, A136 

\bibitem[Manara et al.(2017)]{Manara17a} Manara, C.~F., Testi, L., Herczeg, G.~J., et al.\ 2017, arXiv:1704.02842 


\bibitem[Morbidelli \& Raymond(2016)]{Morby16} Morbidelli, A., \& Raymond, S.~N.\ 2016, Journal of Geophysical Research (Planets), 121, 1962 

\bibitem[Parker et al.(2014)]{Parker14} Parker, R.~J., Wright, N.~J., Goodwin, S.~P., \& Meyer, M.~R.\ 2014, \mnras, 438, 620 

\bibitem[Pfalzner et al.(2005)]{Pfalzner05} Pfalzner, S., Umbreit, S., \& Henning, T.\ 2005, \apj, 629, 526 

\bibitem[Pfalzner et al.(2015)]{Pfalzner15} Pfalzner, S., Vincke, K., \& Xiang, M.\ 2015, \aap, 576, A28 

\bibitem[Robin et al.(2012)]{GUMS} Robin, A.~C., Luri, X., Reyl{\'e}, C., et al.\ 2012, \aap, 543, A100 

\bibitem[Sacco et al.(2015)]{Sacco15} Sacco, G.~G., Jeffries, R.~D., Randich, S., et al.\ 2015, \aap, 574, L7 

\bibitem[Sanchez et al.(2014)]{Sanchez14} Sanchez, N., In{\'e}s G{\'o}mez de Castro, A., et al.\ 2014, \aap, 572, A89 

\bibitem[Thommes et al.(2008)]{Thommes08} Thommes, E.~W., Matsumura, S., \& Rasio, F.~A.\ 2008, Science, 321, 814 

\bibitem[Whittet et al.(1997)]{Whittet97} Whittet, D.~C.~B., Prusti, T., Franco, G.~A.~P., et al.\ 1997, \aap, 327, 1194 

\end{thebibliography}
\end{document}